\documentclass[preprint,showpacs,preprintnumbers,amsmath,amssymb,tightenlines]{revtex4}
\usepackage{amssymb}
\usepackage{amsmath}
\usepackage{graphicx}
\usepackage{epsfig}
\usepackage{subfigure}
\usepackage{amsfonts}
\usepackage{CJK}

\begin{document}

\title{Arbitrary control of coherent dynamics for distant qubits in a quantum
network}
\author{Shi-Biao Zheng$^{1,2}$\thanks{E-mail: sbzheng@pub5.fz.fj.cn},
Chui-Ping Yang$^{1,3,4}$, and Franco Nori$%
^{1,4}$}
\date{\today }

\address{$^1$Advanced Science Institute, Riken, Wako-shi, Saitama 351-0198,
Japan}
\address{$^2$Department of Physics, Fuzhou University, Fuzhou 350002,
China}
\address{$^3$Department of Physics, Hangzhou Normal University,
Hangzhou, Zhejiang 310036, China}
\address{$^4$Department of Physics,
University of Michigan, Ann Arbor, MI 48109, USA}

\begin{abstract}
We show that the coherent coupling of atomic qubits at distant nodes of a
quantum network, composed of several cavities linked by optical fibers, can
be arbitrarily controlled via the selective pairing of Raman transitions.
The adiabatic elimination of the atomic excited states and photonic states
leads to selective qubit-qubit interactions, which would have important
applications in quantum information processing. Quantum gates between any
pair of distant qubits and parallel two-qubit operations on selected qubit
pairs can be implemented through suitable choices of the parameters of the
external fields. The selective pairing of Raman transitions also allows the
generation of spin chains and cluster states without the requirement that
the cavity-fiber coupling be smaller than the detunings of the Raman
transitions.
\end{abstract}
\maketitle

\begin{center}
\textbf{I. INTRODUCTION}
\end{center}

Controlling the coherent dynamics of many-qubit systems lies at the heart of
quantum information. In the standard model of a quantum computer, quantum
information is stored in a quantum register composed of many qubits. The
performance of a certain quantum computational task corresponds to the
control of the unitary coherent evolution of the qubits. Two-qubit quantum
phase gates and multi-qubit entanglement have been achieved experimentally
in various systems. For example, cavity QED with atoms trapped in an optical
cavity can couple atomic qubits via photons (see, e.g., [1]). However, the
practical implementation of quantum computing requires a large number of
qubits, which is extremely difficult to achieve experimentally in a single
cavity. This is due to the fact that the spatial separation between
neighboring qubits decreases as the number of qubits increases, and thus
individual addressing becomes increasingly difficult.

The coherent coupling of separate qubits is of importance for implementing
deterministic long-distance entanglement and large-scale quantum information
processing. The entanglement of distant qubits is an essential ingredient
for testing quantum nonlocality against local-hidden-variable theories [2,3]
and a key resource for quantum communication [4,5]. Furthermore, quantum
logic operations between distant qubits at separate nodes in a network are a
prerequisite for linking several spatially-separated quantum registers to
build a quantum computer. Moreover, the next-nearest-neighbor interaction in
a spin-1/2 chain can be useful for producing cluster states (see, e.g.,
[6,7]), which is the resource for one-way quantum computation.

Recently, schemes have been proposed [8-10] to realize quantum
communication, deterministic entanglement, and phase gates between two
atomic qubits trapped in separate optical cavities, which are coupled by an
optical fiber via coherent dynamics. These previous works [8-10] concentrate
on the simplest case: the system is only composed of two nodes. In order to
implement a distributed quantum computational network with several nodes, it
is necessary to be able to control the coupling of different nodes, exploit
suitable coupling dynamics to perform desired logic operations between any
pair of nodes, and engineer entanglement among these nodes. Unfortunately,
these issues have not been addressed yet.

The controlled dynamics of strongly-interacting many-particle systems is
also of importance in studying quantum phase transitions, which involves
complex collective quantum mechanical behavior. Also, much attention has
been devoted to the ground state entanglement in spin chains near and at the
critical point [11-13], which is responsible for long-range correlations.
Recently, Hartmann \textsl{et al.} [14] have shown that effective spin
lattices can be produced with atoms trapped in an array of microcavities.
The off-resonant Raman transitions between two ground states, induced by the
cavity modes and external fields, lead to spin-spin coupling. In order to
generate a spin chain, in which each qubit is only coupled to its nearest
neighbors, the tunneling rate of photons between neighboring cavities should
be much smaller than the detunings of the Raman transitions so that the
nearest-neighbor coupling dominates the dynamics. Meanwhile, the photon
tunneling rate should be much larger than the cavity decay rate. It is
extremely hard to simultaneously satisfy these two requirements in
experiments.

Here we theoretically show that one can arbitrarily control the coherent
coupling dynamics of multiple atomic qubits at distant nodes of a quantum
network, which is composed of several cavities linked by optical fibers.
This is based on the pairing of off-resonant Raman transitions, through
which the Raman transitions of each qubit can only be coupled to those of
selected qubits to produce the desired qubit-qubit interaction. We present
two applications of this physical mechanism. First, we show that gate
operations between any pair of atomic qubits and selective parallel
two-qubit operations on different qubit pairs can be implemented in the
quantum network without exciting both the atoms and field modes, which could
be useful towards future scalable quantum computing networks. Second, we
show that various spin-1/2 chains can be constructed. As the cavity-fiber
coupling does not need to be smaller than the detunings of the Raman
transitions, much stronger spin-spin couplings can be obtained, offering the
possibility for producing cluster states and observing quantum phenomena in
strongly correlated quantum many-body systems, which were previously not
experimentally accessible.

This paper is organized as follows. In Sec.~2, we study the coherent
coupling dynamics of multiple atomic qubits trapped in separate cavities
linked by optical fibers. We show that, under certain conditions, the atomic
excited states and photonic states can be adiabatically eliminated, and one
can pair off-resonant Raman transitions to produce controlled spin
couplings. In Sec.~3, we present a scheme to implement gate operations
between any pair of atomic qubits and selective parallel two-qubit
operations on different qubit pairs in a quantum network based on controlled
spin couplings. In Sec.~4, we show that spin chains and cluster states can
be generated through pairing off-resonant Raman transitions for neighboring
qubits. In Sec.~5, we address several experimental issues. Conclusions
appear in Sec.~6.

\begin{figure}[tbp]
\includegraphics[bb=81 434 357 544, width=7.6 cm, clip]{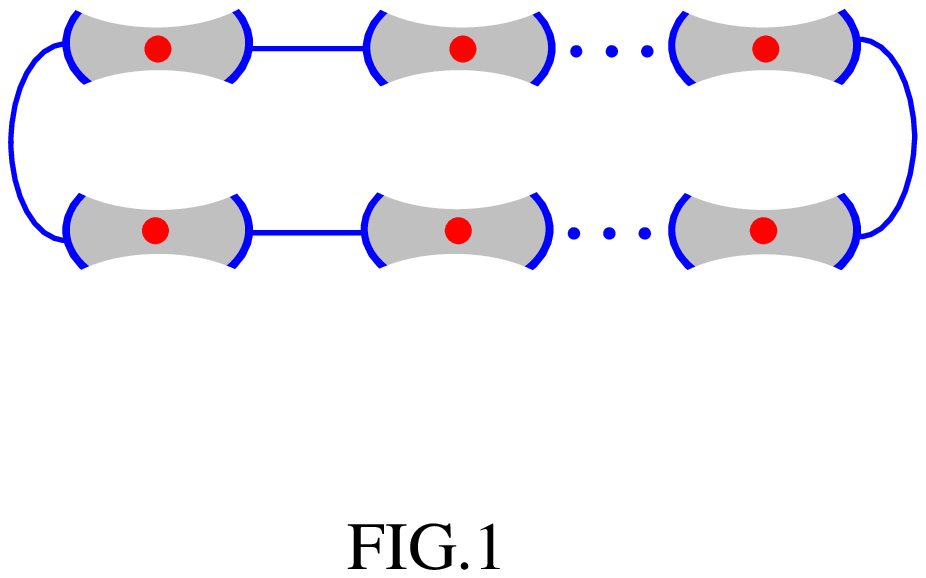} %
\vspace*{-0.08in}
\caption{(color online) Schematic diagram of $n$ distant atoms trapped in
separate coupled cavities, which are connected by short optical fibers.}
\label{fig:1}
\end{figure}

\begin{figure}[tbp]
\includegraphics[bb=150 215 433 724, width=8.6 cm, clip]{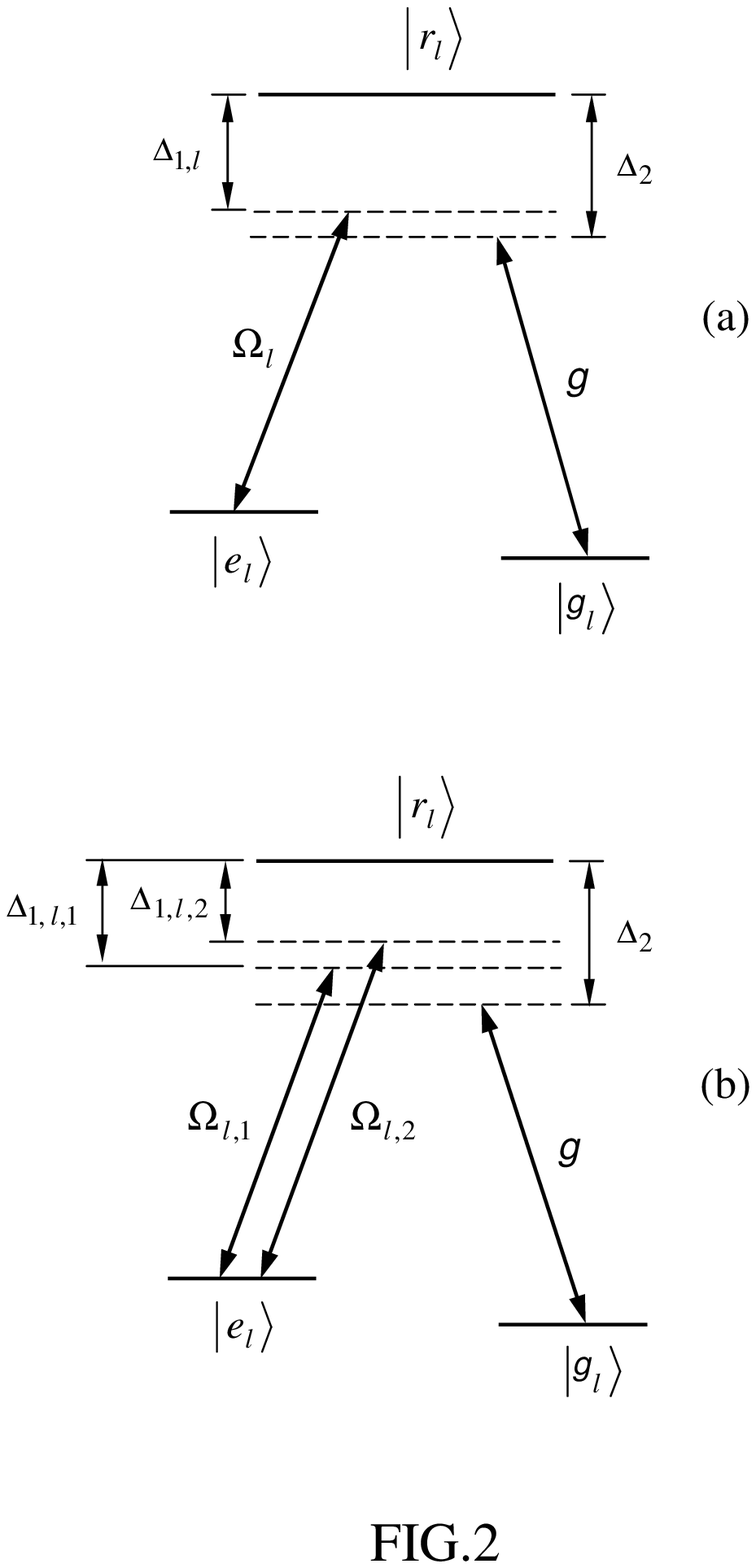} %
\vspace*{-0.08in}
\caption{(a) Atomic level configuration and transitions to implement quantum
gates between any pair of qubits and parallel two-qubit operations. The
transition $\left| e_l\right\rangle \rightarrow \left| r_l\right\rangle $ of
the $l$th atom is driven by a classical laser field with detuning $\Delta
_{1,l}$ and Rabi frequency $\Omega _l$, while the transition $\left|
g_l\right\rangle \rightarrow \left| r_l\right\rangle $ is coupled to the
cavity mode with the coupling constant $g$ and has a detuning $\Delta _2$.
(b) Atomic level configuration and transitions to implement the XY model.
The transition $\left| e_l\right\rangle \rightarrow \left| r_l\right\rangle $
of the $l$th atom is driven by two classical laser fields with detunings $%
\Delta _{1,l,1}$ and $\Delta _{1,l,2}$ and Rabi frequencies $\Omega _{l,1}$
and $\Omega _{l,2}$, while the transition $\left| g_l\right\rangle
\rightarrow \left| r_l\right\rangle $ is coupled to the cavity mode with the
coupling constant $g$ and has a detuning $\Delta _2$.}
\label{fig:2}
\end{figure}

\begin{center}
\textbf{II. CONTROLLED COHERENT COUPLING DYNAMICS}
\end{center}

Let us consider $n$ identical atoms trapped in separated cavities connected
by optical fibers, as shown in Fig. 1. In the short-fiber limit, essentially
only one fiber mode interacts with the cavity modes [9]. We denote $%
\sum_l\equiv \sum_{l=1}^n$, $\sum_m\equiv \sum_{m=1}^n$, and $\sum_k\equiv
\sum_{k=1}^{2n}$. Then the coupling between the cavity modes and fibers are
given by the interaction Hamiltonian
\begin{equation}
H_1=\sum_l\nu b_l\,(a_l^{\dagger }+a_{l+1}^{\dagger })+H.c.,
\end{equation}
where $b_l$ is the annihilation operator for the $l$th fiber mode, $%
a_l^{\dagger }$ is the creation operator for the $l$th cavity mode, and $\nu
$ is the cavity-fiber coupling strength. Here we adopt periodic boundary
conditions, i.e., $b_{n+1}=b_1$, which can be satisfied by linking the first
and the $n$th cavity using another fiber. The atomic level configuration is
shown in Fig. 2(a). Each atom has one excited state $\left| r\right\rangle $
and two ground states $\left| e\right\rangle $ and $\left| g\right\rangle $.
The transition $\left| e_l\right\rangle \rightarrow \left| r_l\right\rangle $
of qubit $l$ is driven by a classical laser field with Rabi frequency $%
\Omega _l$, while the transition $\left| g_l\right\rangle \rightarrow \left|
r_l\right\rangle $ is coupled to the cavity mode with the coupling constant $%
g$. Let us assume that the classical field driving the $l$th atom and the cavity
mode are detuned from the respective transitions by $\Delta _{1,l}$ and $%
\Delta _2$, respectively. In the interaction picture, the Hamiltonian
describing the atom-field interaction is
\begin{equation}
H_2=\sum_l\left( \Omega _le^{i\Delta _{1,l}t}\left| r_l\right\rangle
\left\langle e_l\right| +ga_le^{i\Delta _2t}\left| r_l\right\rangle
\left\langle g_l\right| \right) +H.c..
\end{equation}
Introducing the nonlocal bosonic modes
\begin{eqnarray}
c_l=\frac 1{\sqrt{2n}}\sum_m\left[ e^{-i(2m-1)l\pi /n}a_m+e^{-i2ml\pi
/n}b_m\right] ,
\end{eqnarray}
we can rewrite the Hamiltonians $H_1$ and $H_2$ as
\begin{equation}
H_1=2\nu \sum_k\cos (\pi k/n)c_k^{\dagger }c_k,
\end{equation}
and
\begin{equation}
H_2=\sum_l\left[ \Omega _le^{i\Delta _{1,l}t}\left| r_l\right\rangle
\left\langle e_l\right| +ge^{i\Delta _2t}\sum_ke^{i(2l-1)k\pi /n}c_k\left|
r_l\right\rangle \left\langle g_l\right| \right] +H.c..
\end{equation}
Performing the transformation $\exp (iH_1t),$ we obtain the atom-field
tunable interaction Hamiltonian
\begin{equation}
H_2^{^{\prime }}=\sum_l\left[ \Omega _le^{i\Delta _{1,l}t}\left|
r_l\right\rangle \left\langle e_l\right| +\frac 1{\sqrt{2n}%
}g\sum_ke^{i(2l-1)k\pi /n+i[\Delta _2-2\nu \cos (\pi k/n)]t}c_k\left|
r_l\right\rangle \left\langle g_l\right| \right] +H.c.
\end{equation}
Under the conditions
\[
\Delta _{1,l},\ \left| \Delta _2-2\nu \cos (\pi k/n)\right| \gg \Omega ,\
\frac 1{\sqrt{2n}}g,
\]
the upper level $\left| r_l\right\rangle $ in the Hamiltonian $H_2^{^{\prime
}}$ can be adiabatically eliminated, leading to the couplings between the
two ground states
\begin{eqnarray}
H_{\mathrm{int}} &=&-\sum_l\left\{ \eta _l\left| e_l\right\rangle
\left\langle e_l\right| +\sum_k\left[ \xi _kc_k^{\dagger }c_k\left|
g_l\right\rangle \left\langle g_l\right| \right. \right.  \nonumber \\
&&\ \ +\left. \left. \lambda _{k,l}(c_kS_l^{+}e^{-i(2l-1)k\pi /n}e^{i\delta
_{k,l}t}+H.c.)\right] \right\} ,
\end{eqnarray}
where
\begin{eqnarray*}
\eta _l &=&\Omega _l^2/\Delta _{1,l}, \\
\xi _k &=&g^2\left[ 2n(\Delta _2-2\nu \cos (k\pi /n)\right] ^{-1}, \\
\lambda _{k,l} &=&\frac{\Omega _lg}{2\sqrt{2n}}\left\{ \Delta _1^{-1}+\left[
\Delta _2-2\nu \cos (k\pi /n)\right] ^{-1}\right\} , \\
\delta _{k,l} &=&\Delta _2-2\nu \cos (\pi k/n)-\Delta _{1,l}, \\
S_l^{+} &=&\left| e_l\right\rangle \left\langle g_l\right| ,\text{ }%
S_l^{-}=\left| g_l\right\rangle \left\langle e_l\right| .
\end{eqnarray*}

The Hamiltonian $H_{\mathrm{int}}$ describes multiple off-resonant Raman
transitions for each atom induced by the classical field and the bosonic
modes $c_k$. Under the condition $\delta _{k,l}\gg \lambda _{k,l}$,$\eta _l$%
, $\xi _k$, the bosonic modes do not exchange quanta with the atomic system.
The off-resonant Raman coupling leads to Stark shifts and couplings between
the atoms. Then the effective Hamiltonian becomes
\begin{eqnarray}
H_{\mathrm{eff}} &=&-\sum_l\left\{ \eta _l\left| e_l\right\rangle
\left\langle e_l\right| +\sum_k\left[ \xi _kc_k^{\dagger }c_k\left|
g_l\right\rangle \left\langle g_l\right| +\right. \right.   \nonumber \\
&&\ \ \ \ \ \ \ \ \ \mu _{k,l}(c_k^{\dagger }c_k\left| g_l\right\rangle
\left\langle g_l\right| -c_kc_k^{\dagger }\left| e_l\right\rangle
\left\langle e_l\right| \ )  \nonumber \\
&&\ \ \ \ \ \ \ \ \ \left. +\sum_m\left. (\chi
_{k,l,m}\,S_l^{+}S_m^{-}\,e^{i(\Delta _{1,m}-\Delta _{1,l})t}+H.c.)\right]
\right\} ,
\end{eqnarray}
where $l\neq m$, $\mu _{k,l}=\lambda _{k,l}^2/\delta _{k,l}$, and
\[
\chi _{k,l,m}=\frac 12\lambda _{k,l}\lambda _{k,m}(\delta _{k,l}^{-1}+\delta
_{k,m}^{-1})e^{-2i(l-m)k\pi /n}.
\]
As the quantum number of the bosonic modes conserves during the interaction,
they will remain in the vacuum state if they are initially in the vacuum
state. Then the effective Hamiltonian $H_{\mathrm{eff}}$ reduces to
\begin{equation}
H_{\mathrm{eff}}=\sum_l\left\{ \varepsilon _l\left| e_l\right\rangle
\left\langle e_l\right| +\sum_m\left[ \chi _{l,m}S_l^{+}S_m^{-}e^{i(\Delta
_{1,m}-\Delta _{1,l})t}+H.c.\right] \right\} ,
\end{equation}
where $l\neq m$, $\varepsilon _l=\sum_k\mu _{k,l}-\eta _l$, and $\chi
_{l,m}=\sum_k\chi _{k,l,m}$. The Hamiltonian (9) has the same
form as the Hamiltonian describing the coupling between quantum dots in a single cavity
[15]. However, the coupling between qubits is induced by multiple nonlocal
bosonic modes, while the qubit-qubit coupling in Ref. [15] was induced by a single cavity mode.
Since the \textit{coupling} strength $\chi _{l,m}$ and the \textit{detuning}
$\left( \Delta _{1,m}-\Delta _{1,l}\right) $ can be \textit{controlled} via
the external fields, the effective coupling Hamiltonian $H_{\mathrm{eff}}$
can be used to realize a variety of quantum logic and entanglement
operations between qubits trapped in separated cavities. We note that the
Hamiltonian $H_{\mathrm{eff}}$ can also be obtained in an array of coupled
cavities without using optical fibers [14,16]. We now consider the case when
both the classical field and cavity mode drive the transition $\left|
g\right\rangle \rightarrow \left| r\right\rangle $. Under the above-
mentioned large-detuning conditions, the effective Hamiltonian $H_{\mathrm{%
eff}}$ is given by
\begin{equation}
H_{\mathrm{eff}}=\sum_l\left\{ \varepsilon _l\left| g_l\right\rangle
\left\langle g_l\right| +\sum_m\left[ \chi _{l,m}\left| g_l\right\rangle
\left\langle g_l\right| \otimes \left| g_m\right\rangle \left\langle
g_m\right| e^{i(\Delta _{1,m}-\Delta _{1,l})t}+H.c.\right] \right\} ,
\end{equation}
where $l\neq m$.

\begin{center}
\textbf{III. SELECTIVE GATE OPERATIONS}
\end{center}

Let us now set
\[
\Omega _p=\Omega _q=\Omega ,\;\Omega _l=0\text{ }(l\neq p,q),\;\Delta
_{1,q}=\Delta _{1,p}.
\]
Then we have
\[
\lambda _{k,l}=0\text{ }(l\neq p,q),\chi _{l,m}=0\text{ }(l\neq p,q\text{ or
}m\neq p,q),\varepsilon _p=\varepsilon _q=\varepsilon .
\]
In this case, the coupling Hamiltonian (9) reduces to
\begin{equation}
H_{\mathrm{eff}}=\varepsilon \left( \left| e_p\right\rangle \left\langle
e_p\right| +\left| e_q\right\rangle \left\langle e_q\right| )+(\chi
_{p,q}S_p^{+}S_q^{-}+H.c.\right) .
\end{equation}
This Hamiltonian, describing the selective coupling between qubits $p$ and $%
q $, can be used to perform \textit{entangling} operations between qubits $p$
and $q$. For example, assume that the two atoms are initially in the state $%
\left| e_p\right\rangle \left| g_q\right\rangle $. After an interaction time
$t=\pi /(4\chi _{p,q}),$ the two qubits evolve to the maximally-entangled
state $(\left| e_p\right\rangle \left| g_q\right\rangle -i\left|
g_p\right\rangle \left| e_q\right\rangle )/\sqrt{2}$ [17]. The coherent
dynamics also allows quantum state transfer between the two distant qubits.
Suppose now that qubit $p$ is initially in a superposition of states $\left|
e_p\right\rangle $ and $\left| g_p\right\rangle $, and that qubit $q$ is
initially in the state $\left| g_q\right\rangle $. After an interaction time
$t=\pi /(2\chi _{p,q})$, the initial state of qubit $p$ is transferred to $q$%
.

Selective parallel two-qubit operations can also be implemented. As an
example, suppose that one wants to perform gates on qubit pairs ($p$, $q$)
and ($u$, $v$). Then we drive each of these qubits with a laser field. The
frequencies of these classical fields are suitably adjusted so that $\Delta
_{1,p}=\Delta _{1,q}$, $\Delta _{1,u}=\Delta _{1,v}$, and $\left| \Delta
_{1,p}-\Delta _{1,u}\right| \gg \left| \chi _{\alpha ,\beta }\right| $ ($%
\alpha =p$, $q$ and $\beta =u$,$v$). In this case, qubit $p$ or $q$ is
decoupled to $u$ or $v$ due to the large detunings. Thus qubit $p$ is only
coupled to qubit $q$, and $u$ only coupled to $v$, i.e., the Raman
transitions of qubit $p$ ($u$) is paired with those of qubit $q$ ($v$).
Setting $\Omega _p=\Omega _q$ and $\Omega _u=\Omega _v$ we have $\varepsilon
_p=\varepsilon _q$ and $\varepsilon _u=\varepsilon _v$. The effective
Hamiltonian is now given by
\begin{equation}
H_{\mathrm{eff}}=\varepsilon _p\sum_{s=p,q}\left| e_s\right\rangle
\left\langle e_s\right| +\varepsilon _u\sum_{\mu =u,v}\left| e_\mu
\right\rangle \left\langle e_\mu \right| +\left( \chi
_{p,q}S_p^{+}S_q^{-}+\chi _{u,v}S_u^{+}S_v^{-}+H.c.\right) .
\end{equation}
As the coherent coupling between qubits $p$ and $q$ is not affected by that
between $u$ and $v$, entangling and swap gates on qubit pairs ($p$, $q$) and
($u$, $v$) can be simultaneously performed.

The effective Hamiltonian (10) allows the implementation of controlled phase
gates between any pair of qubits and parallel two-qubit phase gates through
a suitable choice of the Rabi frequencies and detunings of the classical
fields. It should be noted that the selective parallel two-qubit operations
are not restricted to the case when the selected qubit pairs undergo the
same kind of gate transformations. For example, assume now that the
transition $\left| e\right\rangle \leftrightarrow \left| r\right\rangle $ of
each of qubits $p$ and $q$ is driven by a laser field with the detuning $%
\Delta _{1,p}$, while transition $\left| g\right\rangle \leftrightarrow
\left| r\right\rangle $ of each of qubits $u$ and $v$ is driven by a laser
field with the detuning $\Delta _{1,u}$. Under the condition that $\left|
\Delta _{1,p}-\Delta _{1,u}\right| $ is much larger than the respective
Raman couplings, the effective Hamiltonian becomes
\begin{equation}
H_{\mathrm{eff}}=\varepsilon _p\sum_{s=p,q}\left| e_s\right\rangle
\left\langle e_s\right| +\varepsilon _u\sum_{\mu =u,v}\left| e_\mu
\right\rangle \left\langle e_\mu \right| +\left( \chi
_{p,q}S_p^{+}S_q^{-}+H.c.\right) +2\chi _{u,v}\left| g_u\right\rangle
\left\langle g_u\right| \otimes \left| g_v\right\rangle \left\langle
g_v\right| .
\end{equation}
So, in principle, one can simultaneously perform different kinds of gates on
qubit pairs ($p$, $q$) and ($u$, $v$), respectively.

\begin{center}
\textbf{IV. GENERATION OF SPIN CHAINS}
\end{center}

We note that spin chains can also be produced with such a system. We now
assume that the transition $\left| e_l\right\rangle \rightarrow \left|
r_l\right\rangle $ of the $l$th atom is driven by two classical laser
fields, with detunings $\Delta _{1,l,1}$ and $\Delta _{1,l,2}$, and Rabi
frequencies $\Omega _{l,1}$ and $\Omega _{l,2}$, as schematically shown in
Fig.~2(b). The Hamiltonian describing the Raman couplings between the two
ground states now becomes
\begin{eqnarray}
H_{\mathrm{int}} &=&-\sum_{d=1,2}\sum_l\left\{ \eta _{l,d}\left|
e_l\right\rangle \left\langle e_l\right| +\sum_k\left[ \xi
_kc_k^{+}c_k\left| g_l\right\rangle \left\langle g_l\right| \right. \right.
\nonumber \\
&&\ \ \ \ \ \ \ \ \ \ +\left. \left. \left( \lambda
_{k,l,d}c_kS_l^{+}e^{-i(2l-1)k\pi /n+i\delta _{k,l,d}t}+H.c.\right) \right]
\right\} ,
\end{eqnarray}
where
\begin{eqnarray*}
\eta _{l,d} &=&(\Omega _{l,d})^2/\Delta _{1,l,d}, \\
\lambda _{k,l,d} &=&\frac{\Omega _{l,1}g}{2\sqrt{2n}}\{\Delta
_{1,l,d}^{-1}+[\Delta _2-2\nu \cos (k\pi /n)]^{-1}\}, \\
\delta _{k,l,d} &=&\Delta _2-2\nu \cos (2\pi k/2n)-\Delta _{1,l,d}.
\end{eqnarray*}
Under the conditions $\delta _{k,l,d}\gg \lambda _{k,l,d}$, $\eta _{l,d}$, $%
\xi _{k,d}$, the off-resonant Raman coupling for qubit $l$ induced by the $d$%
th ($d=1,2$) classical field, and that for qubit $m$ induced by the $%
d^{^{\prime }}$-th ($d^{^{\prime }}=1,2$) classical field lead to the
two-qubit coupling with coupling strength
\[
\chi _{l,m,d,d^{^{\prime }}}=\sum_k\frac 12\lambda _{k,l,d}\lambda
_{k,m,d^{^{\prime }}}\left( \delta _{k,l,d}^{-1}+\delta _{k,m,d^{^{\prime
}}}^{-1}\right) \,e^{-i2(l-m)k\pi /n}
\]
and detuning
\[
\Lambda _{l,m,d,d^{^{\prime }}}=\Delta _{1,m,d^{^{\prime }}}-\Delta
_{1,l,d}.
\]
The detunings are suitably chosen so that
\begin{eqnarray*}
\Lambda _{l,l-1,1,2} &=&\Lambda _{l,l+1,2,1}=0, \\
\Lambda _{l,m,1,2} &\gg &\chi _{l,m,1,2}\text{ }(m\neq l-1), \\
\Lambda _{l,m,2,1} &\gg &\chi _{l,m,1,2}\text{ }(m\neq l+1).
\end{eqnarray*}
In this case the Raman transition of qubit $l$ induced by the first (second)
classical field is only paired with that of qubit $l-1$ ($l+1$) induced by
the second (first) classical field and thus each qubit is only
resonantly-coupled to its nearest neighbors. The other two-qubit couplings
can be neglected due to large detunings. Under the condition that the field
modes are initially in the vacuum state, the effective Hamiltonian, obtained
from Eq. (14), is now given by
\begin{equation}
H_{\mathrm{eff}}=\sum_l\left[ \varepsilon _l\left| e_l\right\rangle
\left\langle e_l\right| \ \ +\left( \chi
_{l,l+1}S_l^{+}S_{l+1}^{-}+H.c.\right) \right] ,
\end{equation}
where
\begin{eqnarray*}
\chi _{l,l+1} &=&\sum_ke^{i2k\pi /n}\lambda _{k,l,2}\lambda _{k,m,1}\delta
_{k,l,2}^{-1}, \\
\varepsilon _l &=&\sum_{d=1,2}\left( \sum_k\mu _{k,l,d}-\eta _{l,d}\right) ,
\\
&& \\
\mu _{k,l,d} &=&\lambda _{k,l,d}^2/\delta _{k,l,d},\text{ }\eta
_{j,d}=\Omega _{j,d}^2/\Delta _{1,j,d}.
\end{eqnarray*}
We can adjust the Rabi frequencies of the classical fields so that $\chi
_{l,l+1}=\chi _{l+1,l+2}=\chi $. The energy of the level $\left|
e\right\rangle $ can be made identical for all qubits by using the Stark
shift of another nonresonant classical field. In this case, the effective
Hamiltonian corresponds to the XY model.

We now consider the case when the classical fields and cavity mode both
drive the transition $\left| g\right\rangle \rightarrow \left|
r\right\rangle $. After adiabatically eliminating the upper level $\left|
r\right\rangle $, we obtain the Hamiltonian
\begin{eqnarray}
H_{\mathrm{int}} &=&-\sum_{d=1,2}\sum_l\left\{ \eta _{l,d}\left|
g_l\right\rangle \left\langle g_l\right| +\sum_{k=1}\left[ \xi
_kc_k^{\dagger }c_k\left| g_l\right\rangle \left\langle g_l\right| \right.
\right.   \nonumber \\
&&\ \ \ \ \ \ \ \ \ \ \ \left. \left. +\left( \lambda _{k,l,d}c_k\left|
g_l\right\rangle \left\langle g_l\right| e^{-i(2l-1)k\pi /n+i\delta
_{k,l,d}t}+H.c.\right) \right] \right\} .
\end{eqnarray}
As in the XY model, the detunings of the classical fields are suitably
chosen so that the second classical field driving qubit $l$ is only resonant
with the first classical field driving the qubit $\left( l+1\right) $ and
the two corresponding off-resonant Raman transitions are paired. Then the
population operator $\left| g_l\right\rangle \left\langle g_l\right| $ of
qubit $l$ is only coupled to those of its nearest neighbors. Through a
suitable choice of the Rabi frequencies of the driving fields and tuning of
the energy of the level $\left| g_l\right\rangle $, we can obtain from
Eq.~(16) the effective Hamiltonian
\begin{equation}
H_{\mathrm{eff}}=\sum_l\left( \varepsilon \frac{1-\sigma _{z,l}}2+\chi \frac{%
1-\sigma _{z,l}}2\frac{1-\sigma _{z,l+1}}2\right) ,
\end{equation}
where $\sigma _{z,l}=\left| e_l\right\rangle \left\langle e_l\right| -\left|
g_l\right\rangle \left\langle g_l\right| $. After an interaction time $t=\pi
/\chi $, the evolution operator $e^{-iH_{\mathrm{eff}}t}$ plus the single
qubit rotation $\otimes _{l=1}^ne^{i\varepsilon t\left| g_l\right\rangle
\left\langle g_l\right| }$ leads to the cluster state $\frac
1{2^{n/2}}\otimes _{l=1}^n(\left| g_l\right\rangle \sigma _{z,l+1}+\left|
e_l\right\rangle )$, which are the resources for the one-way quantum
computation [6,7].

\begin{center}
\textbf{V. DISCUSSIONS ON EXPERIMENTAL ISSUES}
\end{center}

We now give a brief discussion on the experimental feasibility of the
proposed scheme. Set $n=3$, $\Omega _1=\Omega _3=\nu =g$, $\Omega _2=0$, $%
\Delta _{1,1}=\Delta _{1,3}=16g$, and $\Delta _2=18.5g$. Then we have
\[
\chi _{1,3}=\sum_ke^{(i4k\pi /n)}\frac{\lambda _{k,1}^2}{\delta _{k,1}}%
=8.238\times 10^{-4}g,
\]
and the time needed to complete the entangling operation between qubits 1
and 3 is $t=\pi /(4\chi _{1,3})\simeq 9.53\times 10^2/g$. The probability
that the atoms undergo a transition to the excited state due to the
off-resonant interaction with the classical fields is $p_1\simeq \Omega
_1^2/\Delta _{1,1}^2=3.9\times 10^{-3}$. Meanwhile, the probability that the
field modes are excited due to off-resonant Raman couplings is $p_2\simeq
\sum_{k=1}\lambda _{k,1}^2/\delta _{k,1}^2\simeq 3.1\times 10^{-3}$. Thus
the effective Hamiltonian $H_{\mathrm{eff}}$ is valid. The effective
decoherence rates due to the atomic spontaneous emission and the field decay are $%
\gamma _e=p_1\gamma $ and $\kappa _e=p_2\kappa $, where $\gamma $ and $%
\kappa $ are the decay rates for the atomic excited state and the field
modes, respectively. We here have assumed that the cavity modes and the fiber
modes have the same decay rate. The requirement $\gamma _e$,$\kappa _e\ll
\chi _{1,3}$ means $\gamma ,\kappa \ll 0.2g$. The parameters in the
microsphere cavity QED experiment reported in Ref. [18] are: $g\simeq 2\pi
\times 20$ MHz, $\gamma \simeq 2\pi \times 2.6$ MHz, and $\kappa \simeq 2\pi
\times 7$ MHz. The corresponding cooperativity factor $g^2/2\gamma \kappa $
is too low for the implementation of the qubit coupling. Set $\gamma \sim
\kappa \sim 3\times 10^{-3}g$. This corresponds to a cooperativity factor $%
g^2/2\gamma \kappa \sim 10^5$, which is predicted to be available [19]. Then
the effective decoherent rates are $\gamma _e=1.17\times 10^{-5}g$ and $%
\kappa _e=9.3\times 10^{-6}g$. The corresponding gate fidelity is about $%
F\simeq 1-(\gamma _e+\kappa _e)t\simeq 98\%$. A near-perfect fiber-cavity
coupling with an efficiency larger than 99.9\% can be realized using
fiber-taper coupling to high-$Q$ silica microspheres [20]. The fiber loss at
852 nm wavelength is about 2.2 dB/km [21], which corresponds to the fiber
decay rate $1.52\times 10^5$ Hz, much smaller than the available cavity
decay rate. This implies that the effective decoherence rate due to the
field decay should be smaller than $p_2\kappa $.

\begin{center}
\textbf{V. CONCLUSIONS}
\end{center}

In conclusion, we have theoretically shown that the coherent coupling of
multiple atoms trapped in separated cavities connected by optical fibers can
be arbitrarily controlled through pairing off-resonant Raman transitions of
different atoms. With this physical mechanism, quantum gates between any
pair of qubits and parallel two-qubit operations in the network can be
performed, and various spin chains can be generated. The cavity-fiber
coupling does not need to be smaller than the detunings of the Raman
transitions. For the same coupling to the cavity mode, the effective
spin-spin coupling in our approach exceeds the previous one [14] by at least
one order of magnitude, which is important for the generation of cluster
states and the observation of ground-state entanglement and quantum phase
transitions in quantum many-body systems. An anisotropic spin chain can be
produced through pairing off-resonant balanced Raman transitions between two
ground atomic states [22], in which the counter-rotating terms $c_kS_l^{+}$
and $c_k^{\dagger }S_l^{+}$ are involved.

\begin{center}
\textbf{ACKNOWLEDGEMENTS}
\end{center}

We acknowledge partial support from the Laboratory of Physical Sciences,
National Security Agency, Army Research Office, National Science Foundation
Under Grant. No. 0726909, DARPA, JSPS-RFBR under Grant No. 09-02--92114,
Grant-in-Aid for Scientific Research (S), MEXT Kakenhi on Quantum
Cybernetics, and FIRST (Funding Program for Innovative R\&D on S\&T). S.B.
Zheng acknowledges support from the National Natural Science Foundation of
China under Grant No. 10674025, the Doctoral Foundation of the Ministry of
Education of China under Grant No. 20070386002, and funds from the State Key
Laboratory Breeding Base of Photocatalysis, Fuzhou University. C.P. Yang
acknowledges partial support from the National Natural Science Foundation of
China under Grant No. 11074062, the Natural Science Foundation of Zhejiang
Province under Grant No. Y6100098, and the funds from Hangzhou Normal University.

\end{document}